\documentclass[prb,preprint,showpacs,superscriptaddress,amssymb]{revtex4}
\usepackage{latexsym}
\usepackage[dvips]{graphicx}
\usepackage{bm}
\usepackage{epsf}
\usepackage{epsfig}
\usepackage{dcolumn}
\usepackage{amsmath}
\newcommand{\barra}[1]{\,\overline{{\mathrm{#1}}}\,}
\begin{document}
\title{Quantum-size effects on chemisorption properties: CO on 
Cu ultrathin films}

\author{L. Mouketo} 
\affiliation{Groupe de Simulations Num\'eriques en Magn\'etisme et Catalyse,
D\'epartement de Physique, Universit\'e Marien NGouabi, BP 69, Brazzaville,
Congo}
\affiliation{Centre for Atomic Molecular Physics and Quantum Optics (CEPAMOQ),
University of Douala, P.O. Box. 8580 Douala, Cameroon} 
\affiliation{The Abdus Salam International Centre for Theoretical Physics,
Trieste 34151, Italy}
\author{N. Binggeli}
\affiliation{The Abdus Salam International Centre for Theoretical Physics,
Trieste 34151, Italy}
\affiliation{IOM-CNR DEMOCRITOS National Simulation Center, Trieste,
Italy}
\author{B. M'Passi-Mabiala}
\affiliation{Groupe de Simulations Num\'eriques en Magn\'etisme et Catalyse,
D\'epartement de Physique, Universit\'e Marien NGouabi, BP 69, Brazzaville,
Congo}

\date{\today}
\begin{abstract}
We address, by means of ab-initio calculations, the origin of the correlation 
that has been observed experimentally between the chemisorption energy of CO 
on nanoscale Cu(001) supported films and quantum-size effects.  
The calculated chemisorption energy shows systematic oscillations, as a 
function of film thickness, with a periodicity corresponding to that of 
quantum-well states at $\bar{\Gamma} $ crossing the Fermi energy. We explain 
this trend based on the oscillations, with film thickness, of the decay 
length on the vacuum side of the quantum-well states at the Fermi energy.  
Contrary to previous suggestions, we find that the actual oscillations   
with film thickness of the density of states per atom of the film at 
the Fermi energy cannot account for the observed trend in the chemisorption  
energy.

\end{abstract}

\maketitle

\section{Introduction}

Understanding and controlling the structure-size dependence of the chemical
activity in materials with reduced dimensionality is a longstanding target of 
surface science, whose practical significance stems from potential applications 
in catalysis, gas sensing, and anticorrosion. A recent exciting development in 
this area is the observation, in metal films a few atomic layers thick, of a 
correlation between quantum-size effects and properties such as 
chemisorption\cite{Danese} and surface oxidation.\cite{Aballe,Aballe2,Xucun,Zhen} 
   
Danese \textit{et al.},\cite{Danese} in particular, have observed experimentally 
systematic oscillations in the desorption temperature of CO adsorbates on 
ultrathin epitaxial Cu(001) films on fcc-Fe(001), as a function of film thickness. 
Such Fe-supported films give rise experimentally to well resolved series of 
quantum-well states (QWS's).\cite{Danese,Ortega} The largest CO desorption 
temperatures were found to coincide with Cu film thicknesses at which a 
QWS crosses the Fermi energy, $E_F$, in inverse photo-emission (IPE) 
spectra taken at normal incidence.\cite{Danese} 

Typically, modifications in chemisorption properties of metal surfaces are 
achieved by monolayer or submonolayer deposition of a second metal, forming a 
surface alloy, and exploiting related local surface mechanisms such as site 
blocking or charge exchanges.\cite{Campbell,Rodriguez}  
Using thin-film QWS's instead is an interesting new approach to tailor surface 
chemisorption properties. In spite of its interest, however, the precise 
mechanism and key parameter behind the QWS-related modifications of the 
chemisorption properties are not yet fully understood.

Very recently, quantum-size effects have also been reported in the initial 
oxidation rate of ultrathin Mg, Al, and Pb supported 
films.\cite{Aballe,Aballe2,Xucun,Zhen} The effect was 
initially suggested to be due to periodic oscillations, with film thickness,  
in the magnitude of the density of states (DOS) of the films at the Fermi 
energy.\cite{Aballe,Xucun,Zhen,Hellman05} The DOS per atom of the film is 
a parameter which is often invoked in model reactivity theories,\cite{Wilke96} 
and which may account for the observed reactivity trends of the Pb 
films\cite{Xucun,Zhen} --- although other parameters have also been 
invoked.\cite{Hu,Liu}
In the case of the Mg films, however, it was observed later on that the actual 
variation in the DOS per atom of the film at $E_F$ could not 
simply account for the order-of-magnitude change in the initial oxidation 
rate of the films.\cite{Binggeli2} Moreover, for the Al films, significant 
differences were observed between the DOS and reactivity trends as a function of 
film thickness.\cite{Aballe2}

In the cases of the Mg and Al films, for which the precursor adsorption mode of the 
O$_2$ molecule on the surface should be physisorption, the key 
parameter responsible for the changes in the initial oxidation rate was then proposed  
to be the decay length in vacuum, $\lambda$, of the electronic local density of states 
of the film at the Fermi energy.\cite{Binggeli2,Binggeli} Modifications in 
$\lambda$ can be expected to have a direct exponential influence on the electron 
transfer rate by resonant tunneling, which is believed to control the initial sticking 
of the O$_2$ molecules on such surfaces.\cite{Hellman03} 
The same argument, however, based on tunneling processes, does not hold in  
the case of CO on Cu(001), where the molecule is clearly chemisorbed and the 
modulated value is the chemisorption energy. In this case, the key factor responsible 
for the observed changes is still an open issue. 

In this work, we investigate by means of {\it ab initio} calculations, the 
correlation between the chemisorption energy of CO molecules on Cu(001) 
films and quantum-size effects. The observed systematic oscillations in the 
chemisorption energy can be understood in terms of periodic modifications, 
with film thickness, in the decay length of the QWS's at the Fermi level.  
The paper is organized as follows. In Section~\ref{sec:Method}, we briefly present the 
calculation method and parameters used. In Section~\ref{sec:QWS}, we examine the 
electronic spectra of free-stranding and fcc-Fe-supported Cu(001) films with 
thicknesses in the range 3-12 monolayers (ML). In Section~\ref{sec:Chem}, we 
examine the CO chemisorption energy on the corresponding films. 
The trends and microscopic origin of the correlation 
between electron quantum-size effects and chemisorption energies are analyzed in 
Section~\ref{sec:Analysis}.  Our conclusions are summarized in 
Section~\ref{sec:Summary}.

\section{Computational details}\label{sec:Method}

The calculations were carried out within the generalized gradient approximation 
(GGA) to density-functional theory (DFT), using the Perdew-Burke-Ernzerhof (PBE) 
exchange-correlation functional.\cite{pbe}  
We employed the pseudopotential-plane-wave method, as implemented in  the 
PWscf code of the QUANTUM-ESPRESSO distribution.\cite{Qesp}  
For carbon, oxygen, iron, and copper, we used the 
Rabe-Rappe-Kaxiras-Joannopoulos ultrasoft pseudopotentials of the PWscf
library. The semicore Cu 3d states were treated as valence states.  
The nonlinear core correction\cite{nnlcc} to the exchange-correlation 
potential was used for Cu and Fe.

In the experimental study by Danese \textit{et al.},\cite{Danese} the 
Cu films were grown on a closely latticed matched template formed by 5-ML 
of fcc Fe(001) deposited on a Cu(001) substrate. 
In our study, we considered both free-standing and fcc-Fe-supported Cu(001) 
films, with/without the CO adsorbates.  All calculations involving 
the Cu films on Fe were spin-polarized calculations. 
The systems were modeled using slab geometries in supercells.  
The length of the supercell was set to 85 \AA\@ in all cases. 
A kinetic-energy cutoff of 30 Ry was used for the plane-wave expansion of the 
Kohn-Sham orbitals.   

For the supported Cu films, we considered slabs containing 5 Fe(001) 
ML plus 3 to 12 Cu(001) ML deposited on one side. The Cu theoretical lattice 
constant of 3.68 \AA\@ was used to construct the slabs (the experimental 
value is 3.61 \AA); we neglected the effect of the small lattice 
mismatch between fcc-Fe and Cu.\cite{NoteFe}  
To evaluate the work functions of the Cu(001) supported films, we employed 
larger, symmetric slabs of $n$-ML Cu/5-ML Fe/$n$-ML Cu ($3 \leq n \leq 12$).   
The smallest vacuum region used was 31 \AA, corresponding to a 29-ML slab. 
For the CO covered films, we used asymmetric slabs with CO adsorbed on one 
side of the slab. We considered the cases of the Cu(001)-($1\times1$)-CO 
and Cu(001)-c($2\times2$)-CO surfaces, corresponding to CO coverages 
$\Theta = 1$ ML and  $\Theta = 0.5$  ML, respectively. Experimentally, the 
desorption measurements in Ref. \onlinecite{Danese} were performed for the 
Cu(001)-c($2\times2$)-CO surface.  

The adsorption site of the CO molecule is known from experiment to be the 
on-top site.\cite{exp} This is in agreement with our PBE-GGA calculations, and 
consistent with the results of previous GGA calculations using the same type 
of pseudopotentials.\cite{Favot01} 
We note, however, that although our GGA calculations yield the correct 
adsorption site for CO on Cu(001), this is not so in general within 
GGA for other transition metal surfaces - where the underestimation 
of the CO gap gives rise to a different site preference with respect to 
experiment.\cite{Kresse03,Mason04} We therefore also checked 
that changing the CO gap (with a GGA + U scheme, and U$_{CO} = 0.75$ eV, 
as in Ref. \onlinecite{Kresse03}) does not affect (within 0.1 meV) the 
calculated variations of the chemisorption energy with film 
thickness.\cite{NoteDFT}  

The CO molecule is vertical on Cu(001), the experimental  
Cu-CO bonding distance is $d$(Cu-C)$ = 1.92$ \AA\@ and the 
C-O distance is $d$(C-O)$ = 1.13$ \AA.\cite{exp} 
From calculations for the Cu(001)-c($2\times2$)-CO surface, using a 12-ML 
thick Cu film, we find $d$(Cu-C)$ = 1.9$ \AA\@ and $d$(C-O)$ = 1.15$ \AA, in 
good agreement with the experimental values\cite{exp} and with previous GGA 
results for thinner Cu films.\cite{Favot01} We kept these distances fixed in 
the remaining part of our study.  We also elected to use the same frozen Cu 
(Cu/Fe) slab geometry for the CO covered and uncovered Cu films, as our main 
purpose here is to understand the correlation between the CO-Cu(001) 
chemisorption energy and quantum-size effects, which are of electronic origin. 

We evaluated the CO chemisorption energy as: 
\begin{equation}
E_{ch}(CO) = - ( E_{slab+CO} - E_{slab} -  E_{CO} ) 
\end{equation}
where $ E_{slab} $ and $ E_{slab+CO} $ are the total energies 
of the slab without and with the adsorbed  carbon monoxyde, respectively, and 
$E_{CO} $ is the total energy of the CO molecule in the gas phase. 
For the calculation of the isolated carbon monoxide molecule, we have 
used a cubic cell with a parameter of $ a= 15$ \AA. The corresponding equilibrium  
C-O distance was found to be d(C-O)$ = 1.14$ \AA. 

In the self-consistent calculations for the ($1\times1$) [c($2\times2$)] 
surfaces, the integrations over the Brillouin zone were performed using a $24 
\times 24\times 1$  [$18 \times 18 \times 1$]  k-point grid centered at 
$\Gamma$.  A Gaussian smearing of the electronic levels of 0.01 Ry was used 
to determine the Fermi energy. The relative values of the chemisorption energy 
were found to be converged to within  $\sim 1$ meV with respect to the 
kinetic-energy cutoff, k-point grid, and vacuum size.  
For the calculations of the DOS of the film at the Fermi energy, we   
increased the k-point grid to ($ 40\times40\times1$), in order to ensure a 
numerical accuracy of 0.0004 eV$^{-1}$ on the DOS value per atom and spin. 
To calculate the local density of states at $E_F $, used to 
evaluate the decay length $\lambda$, we employed a ($ 48\times48\times1$) 
k-point grid and a Gaussian smearing of 0.005 Ry. We determined 
$ \lambda $ from a fit, assuming an exponential decay of the local density of 
states at distances beyond $\sim 2.4$~\AA\ from the outermost atomic plane. 
The numerical accuracy on $\lambda$ was estimated as 0.002 \AA. 

\section{Results and discussion}

\subsection{Quantum-well states of the free-standing and supported Cu(001) films}\label{sec:QWS}

\begin{figure}[h]
\vspace*{1.0cm}
\begin{center}
\includegraphics[width=0.67\textwidth]{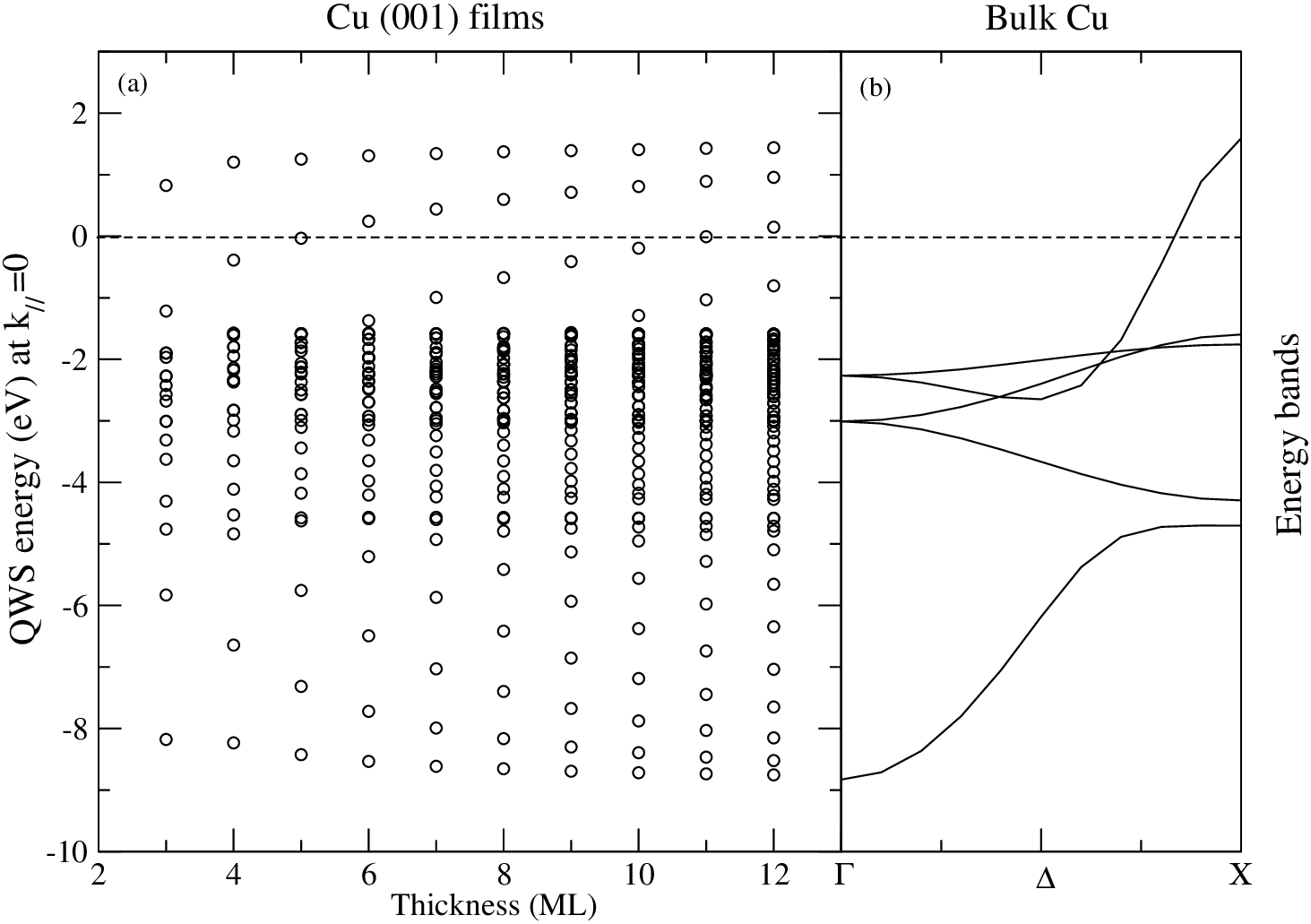}
\caption{\label{qws} (a) Calculated energies of the quantum-well states at 
$\bar{\Gamma} $ in the free-standing Cu(001) films as a function of film 
thickness. (b) Dispersion of the Cu bulk bands along the [001] 
direction (i.e., the $ \Gamma -$ X direction of the bulk Brillouin zone), 
perpendicular to the surface of the films. The zero of energy corresponds 
to the Fermi level. }
\end{center}
\end{figure}

In Fig.\ref{qws}(a), we show the calculated energy levels of the QWS's with 
wavevector $ k_{||}= 0 $ for the free-standing Cu(001) films, as a function 
of film thickness. The unoccupied levels at  $\bar{\Gamma} $ are the 
states probed in normal-incidence IPE experiments.\cite{Danese}  
In Fig.\ref{qws}(b), we also show the energy bands of the Cu bulk 
states with wavevectors perpendicular to the films, i.e., along the $ \Gamma - $
X line ($\Delta$ direction) of the bulk Brillouin zone. 
The QWS's of the films at $\bar{\Gamma} $ originate from these bulk states. 

The QWS's of interest near the Fermi level, in 
Fig.\ref{qws}(a), with energies in the range [$E_{F}-1.5$ eV,  
$E_{F}+1.5$ eV], derive from the upper part of the bulk Cu $4sp$ band,  
in Fig.\ref{qws}(b), which crosses the Fermi energy and has its maximum at 
the X point. The spectrum in the energy range [$E_{F}-5$ eV, $E_{F}-1.5$ eV]   
is dominated by the  more localized Cu $3d$-band states.  The 
lower part of the bulk Cu $4sp$ band (below $E_{F}-5$ eV) produces another 
series of QWS's, in Fig.\ref{qws}(a), located in the energy range [$E_{F}-9$ 
eV, $E_{F}-5$ eV]. 

\begin{figure}[h]
\vspace*{1.0cm}
\begin{center}
\includegraphics[width=0.67\textwidth]{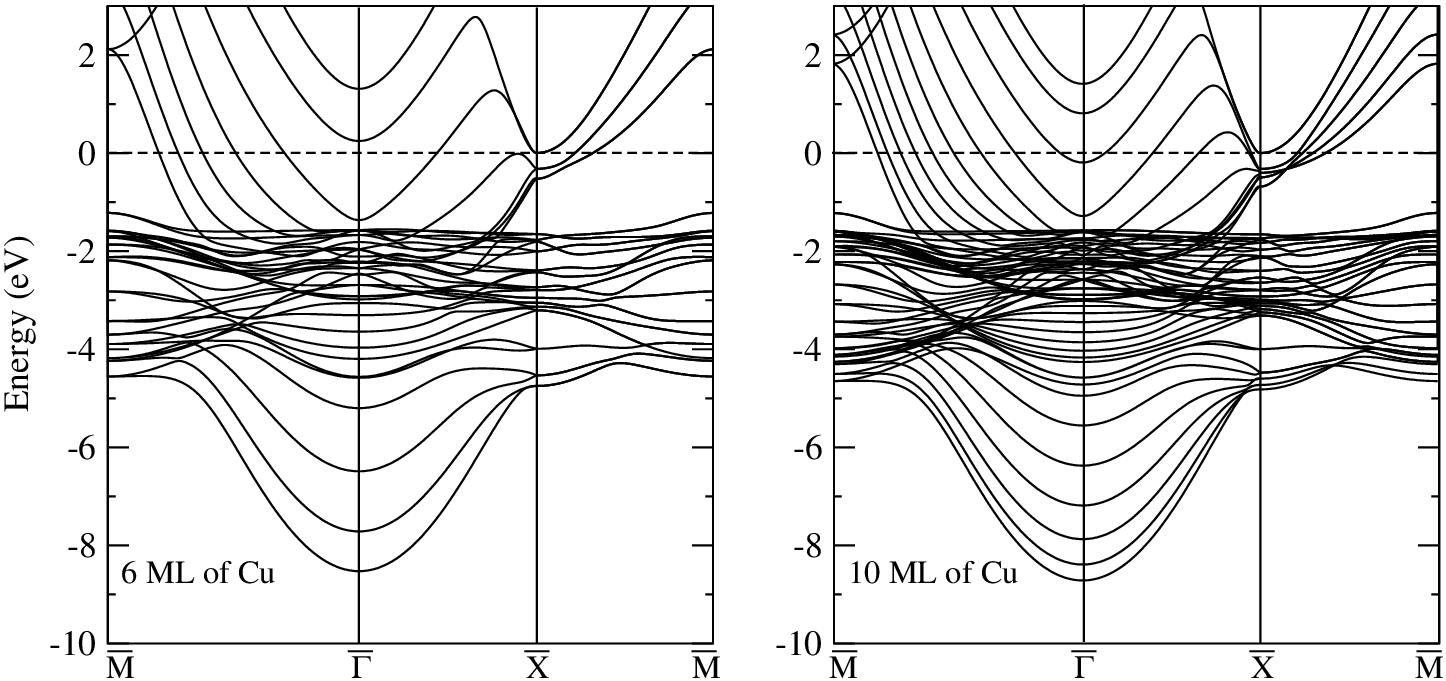}
\caption{\label{band10} Band structure of the 6-ML (left panel) 
and 10-ML (right panel) free-standing Cu(001) films along the high-symmetry lines of 
the surface Brillouin zone. The zero of energy corresponds to the Fermi level.}
\end{center}
\end{figure}

In Fig. \ref{band10}, we display the band structures of the 6-ML and  10-ML 
free-standing Cu(001) films along the high-symmetry lines of the surface 
Brillouin zone. The low-dispersion $3d$ bands can be easily recognized 
in the energy range [$E_{F}-5$ eV, $E_{F}-1.5$ eV]. We also note that the two 
series of quantum-well levels at $\bar{\Gamma} $, originating from the 
top and bottom part of bulk Cu $4sp$ band along the $\Delta$ direction, 
give rise to two characteristic series of parabolic-like subbands, 
in Fig. \ref{band10}, near the surface Brillouin-zone center. 

In Fig.\ref{qws}(a), one can observe that the energies, near $E_{F}$, of the 
QWS's of the film at $ \bar{\Gamma} $ increase with increasing film thickness 
and cross the Fermi energy at 5 ML and at 11 ML. This is in good agreement with the 
IPE measurements, where the crossing occurs at 5 ML and at 10-11 ML.\cite{Danese} 
The trend of increasing energy with increasing film thickness of the $\bar{\Gamma}$ 
QWS's (which is opposite to the trend of the QWS's near $E_{F}$ in 
the Mg(0001), Al(111), and Pb(111) films\cite{Aballe,Aballe2,Xucun}) 
results from the negative effective mass of the $4sp$ band of bulk Cu, at the 
X point, in the $\Delta$-direction. The top of the Cu $4sp$ bulk band, at 
about 1.5 eV above  $E_{F}$ in Fig.\ref{qws}(b), corresponds to the bottom of 
the ``inverted'' 1-D quantum well which determines the energies, near $E_{F}$, 
of the QWS's of the film at $ \bar{\Gamma}$.  With increasing film thickness, 
the quantum-well levels become increasingly closer to the bottom of the 
inverted quantum well at $\sim 1.5$ eV above  $E_{F}$. 

The periodicity of the crossing of $E_F$ by $\bar{\Gamma}$ QWS's can be derived 
from the Bohr-Sommerfeld rule,\cite{Paggel99} considering the parabolic-like 
part of the bulk Cu $4sp$ band in Fig.\ref{qws}(b) and its intersection  
with $E_{F}$. The empty section of the Cu $4sp$ bulk band runs over 17 \% of 
the Brillouin-zone  X $- \Gamma$ line, and therefore every $1/0.17=5.8$ ML 
a quantum-well level of the film at $ \bar{\Gamma} $ should cross the Fermi 
energy, according to the Bohr-Sommerfeld rule.\cite{Paggel99} This is in good 
agreement with the crossing periodicity of 6 ML we find in Fig.\ref{qws}(a), and 
consistent with the 5-6 ML crossing periodicity observed in the IPE 
spectra.\cite{Danese}  Our results are also in good agreement 
with previous DFT calculations for the $ \bar{\Gamma} $-state spectra 
of the free-standing Cu(001) films and their analysis.\cite{Bosun,Han} 

\begin{figure}[h]
\vspace*{1.0cm}
\begin{center}
\includegraphics[width=0.6\textwidth]{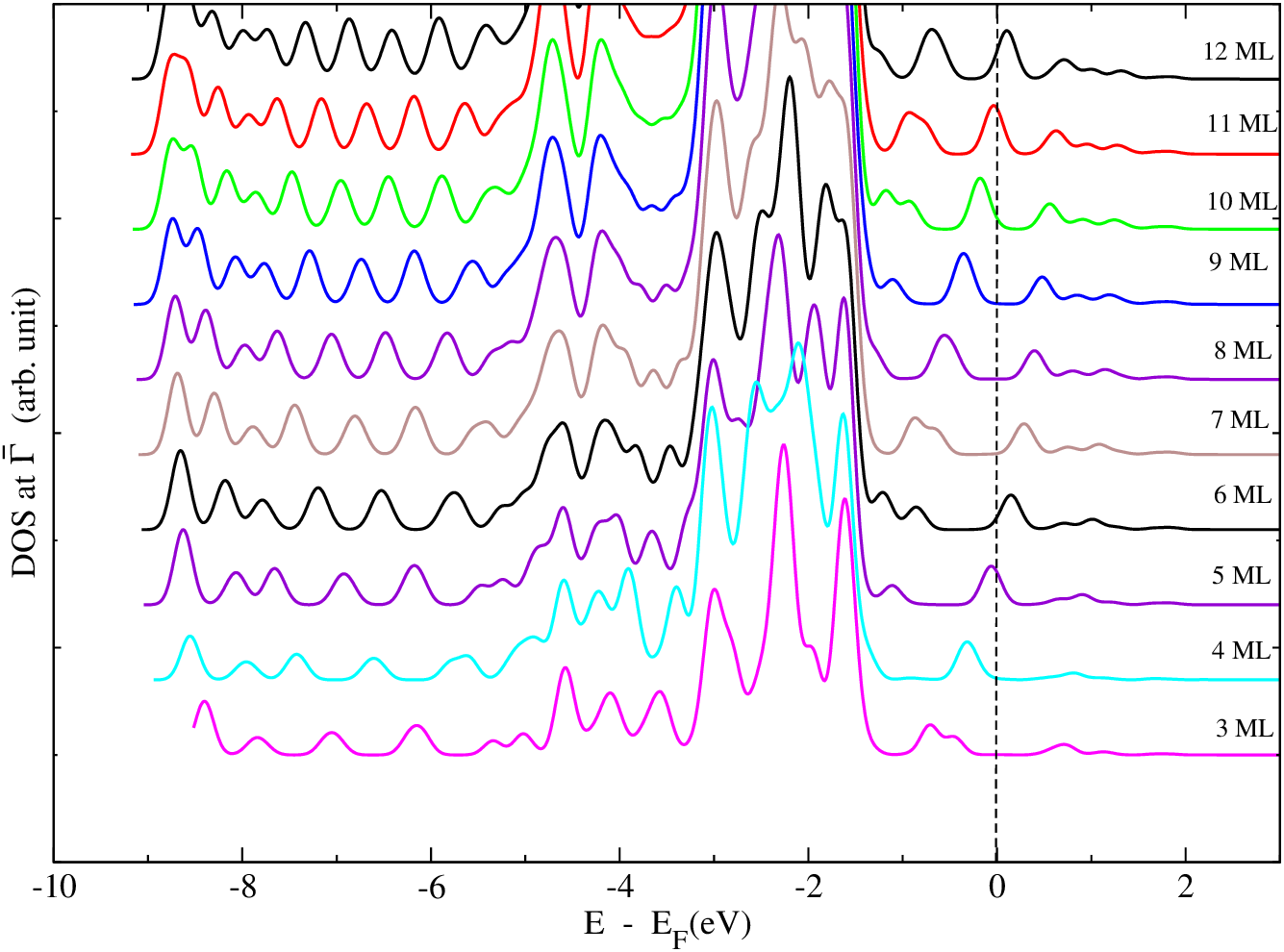}
\caption{\label{supqws} Partial densities of states of the Cu(001) films at
$ \bar{\Gamma} $, for the Fe-supported Cu films. The film thickness increases from 
3 to 12 atomic layers (bottom to top curve).}
\end{center}
\end{figure}

As the experimental Cu films in Ref.~\onlinecite{Danese} were grown on fcc Fe(001), 
we also investigated the effect of the Fe substrate on the QWS spectra. 
In Fig. \ref{supqws}, we show the calculated partial density of states at 
$ \bar{\Gamma}$ of the Cu(001) films on the fcc-Fe(001) substrate, as a function 
of film thickness.\cite{NoteRelax} The partial density of states of the film was 
obtained by summing the spin-up and spin-down atomic-projected density of states 
at $ \bar{\Gamma}$ of the Cu atoms in the film. The results in Fig. \ref{supqws} 
show that the sequence of thicknesses at which the QWS peak maxima cross $E_{F}$ 
remains exactly the same as for the free-standing films, namely 5 and 11 ML. 

\subsection{Quantum-size effects on the CO chemisorption energy}\label{sec:Chem}

\begin{figure}[h]
\vspace*{1.0cm}
\begin{center}
\includegraphics[width=0.75\textwidth]{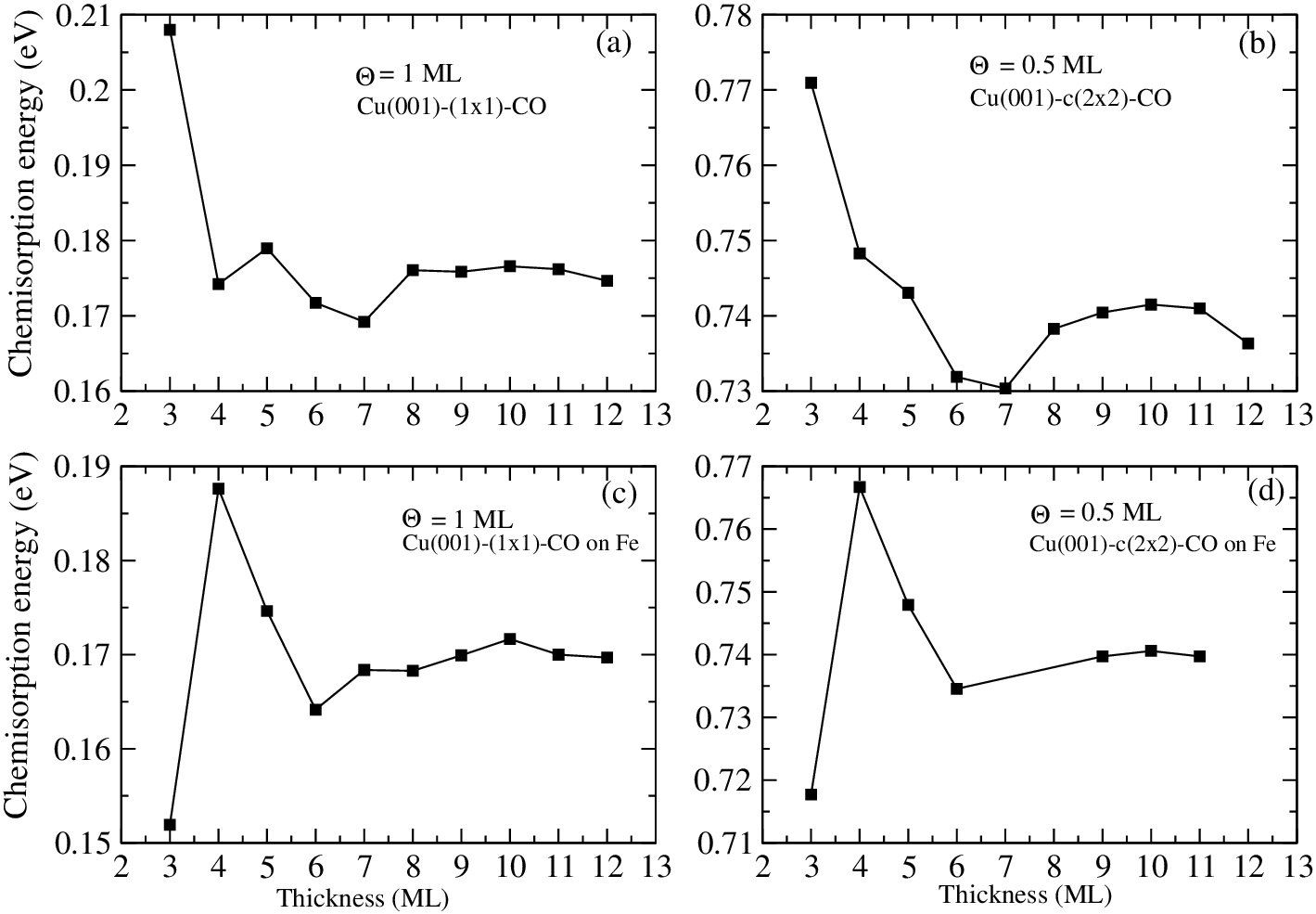}
\caption{\label{ads1} Calculated chemisorption energy of the CO molecule for the 
free-standing (upper panels) and Fe-supported (lower panels) Cu(001) films, as 
a function of film thickness. Left panels show the results for the 
Cu(001)-($1\times1$)-CO surface and right panels for the 
Cu(001)-c($2\times2$)-CO surface, corresponding to CO coverages of 1 ML and 
0.5 ML, respectively. }
\end{center}
\end{figure}

The calculated CO chemisorption energies of the free-standing and Fe-supported 
Cu(001) films are shown in Fig. \ref{ads1}, for the two CO coverages  
$ \Theta = 1$ ML and $\Theta = 0.5$ ML. 
The behavior as a function of film thickness is rather similar for the two 
CO coverages. 
For the free-standing films (and for both $ \Theta = 1$ ML and $\Theta = 0.5$ ML), 
the chemisorption energy is largest at 3 ML, and displays a local minimum at 7 ML 
and a local maximum at 10 ML; for $ \Theta = 1$ ML, in addition, a local minimum 
(maximum) is present at 4 (5) ML. In the case of the Fe-supported films, 
both $\Theta$ coverages give rise to the same trends: local maxima are found 
in the chemisorption energy at 4 and 10 ML, and a local minimum is present 
at 6 ML. 

The presence of the Fe substrate modifies thus the trends of 
the chemisorption energy at low film thickness (below 8 ML), giving rise to 
a pronounced maximum at 4 ML and also shifting the local minimum from 7 ML (in the 
free-standing case) to 6 ML (for the supported films). At larger film thicknesses, 
instead, a local maximum is found at 10 ML in all cases. The presence of the 
substrate gives rise to a systematic oscillatory behavior in the 
chemisorption energy, which improves the agreement with the experimental 
trend.\cite{Danese}  Experimentally, the local maxima in the chemisorption 
energy occur at 5 and 10-11 ML, and a local minimum is observed at 6 ML.\cite{Danese}  

For the supported films, the oscillatory behavior of the chemisorption 
energy shows a correlation with the periodic crossing of $E_F$ by a QWS at 
$ \bar{\Gamma}$. In fact, the theoretical maxima in the chemisorption 
[at 4 and 10 ML, in Fig. \ref{ads1} (c) and (d)] are found to 
occur systematically 1 ML before the crossing of $E_F$ by a QWS 
(at 5 and 11 ML, in Fig. \ref{supqws}). It may appear surprising, 
at first sight, that the presence of the substrate restores a trend which one 
would expect to be intrinsically related to the properties of the film. 
It should not be forgotten, however, that effects other than those induced by 
the QWS near $E_F$ can be expected to have a major influence on the 
chemisorption energy for free-standing films only a few 
monolayers thick. Indeed, the presence of a free Cu surface with broken bounds 
a few Cu ML away from the CO covered surface may be expected to strongly 
modify (enhance) the strength of the CO-Cu bound. 

We find that the chemisorption energy for $ \Theta = 0.5 $ ML is considerably 
larger than that for $ \Theta = 1 $ ML (by $\sim 0.56$ eV, at large film thickness). 
This is consistent with the experimental trend, and an increased CO-CO repulsion at 
coverages larger than 0.5 ML.\cite{Tracy,Mason06,NoteLiu} 
The chemisorption energy for the Cu(001)-c($2\times2$) surface ($\sim 0.74$ eV) 
is consistent with previous GGA values,\cite{Favot01} and somewhat larger 
than the experimental value (0.57 eV).\cite{Tracy}

\subsection{Discussion and microscopic interpretation}\label{sec:Analysis}

In order to better understand the mechanism and identify the key parameter 
responsible for the oscillations in the chemisorption energy, we have 
investigated the behavior with film thickness of (i) the DOS per atom of the 
films at the Fermi energy, (ii) the work function of the films, and (iii) the 
decay length in vacuum ($\lambda$) of the electronic local density of states 
of the films at $E_F$. Both the DOS and $\lambda$ have been invoked as 
possible key factor responsible for the oscillations in the oxidation rate of 
ultrathin metal films, while the work function is a parameter, known to exhibit 
quantum-size effects,\cite{Schulte,Paggel02} which has also been been shown to 
be important when discussing  reactivity, e.g., in relation with catalytic 
promotion.\cite{Imbihl} 

\begin{figure}[h]
\vspace*{1.0cm}
\begin{center}
\includegraphics[width=0.57\textwidth]{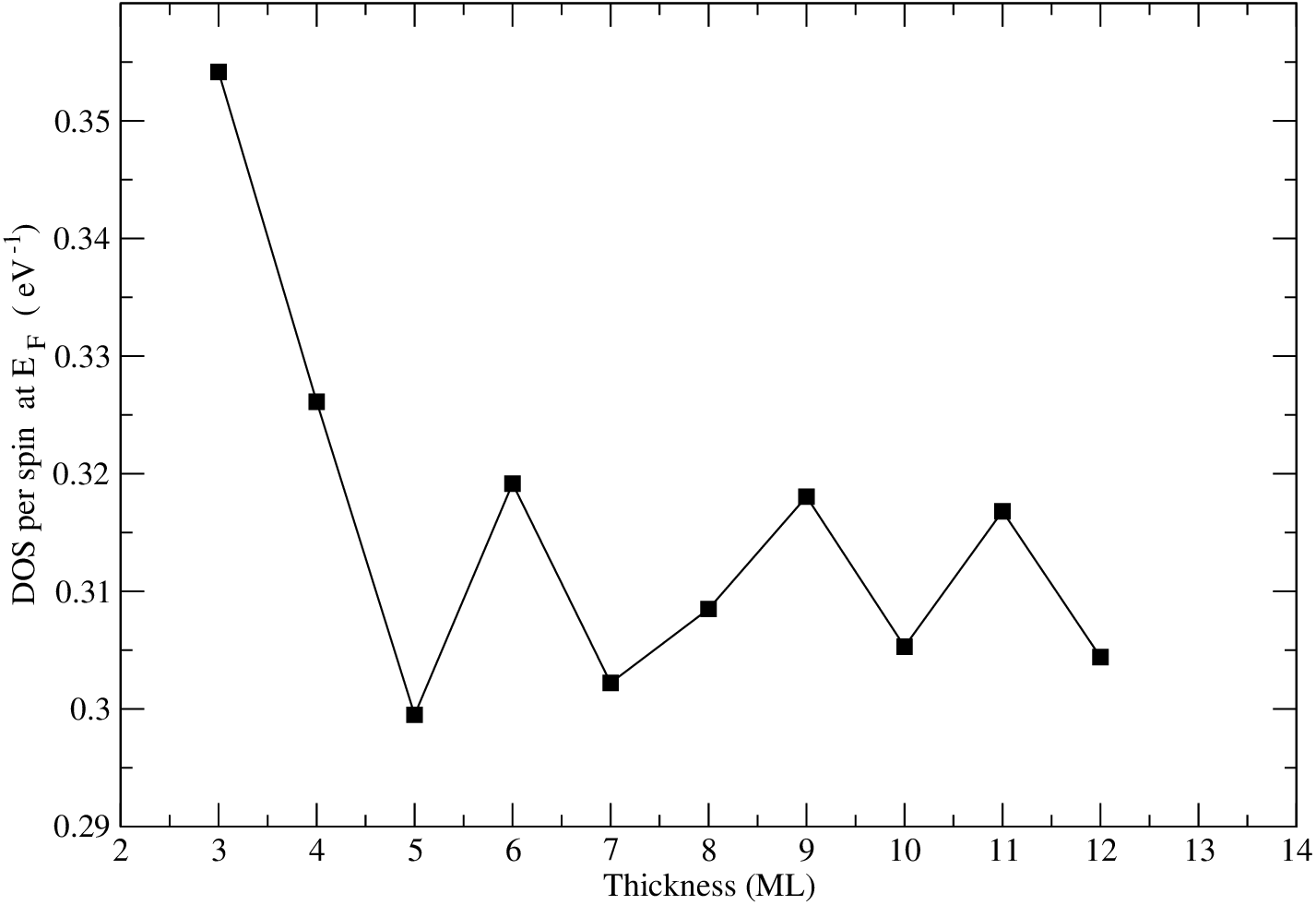}
\caption{\label{dosEF} Calculated density of states per atom at the Fermi energy for 
the free-standing Cu(001) films, as a function of film thickness.}
\end{center}
\end{figure}

In Fig. \ref{dosEF}, we display the calculated DOS per atom at $E_F$ of the 
free-standing Cu(001) films, as a function of film thickness.  
The DOS clearly displays short-period (2 to 3 ML) oscillations, which neither  
follow the periodic crossing between the QWS at $ \bar{\Gamma} $ and $E_F$ 
(in Fig. \ref{supqws}) nor show a correlation with the 
oscillations in the chemisorption energy of the supported films (in 
Fig. \ref{ads1}). We note that the results in Fig. \ref{dosEF} 
are consistent with the calculated DOS($ E_{F}$) behavior reported for 
free-standing films with thicknesses larger than 7 ML in 
Ref.~\onlinecite{Bosun}. The short-period oscillations of the DOS($ E_{F}$) 
can be related to a beating effect due to a supersposition of short and long 
Fermi-wavelength oscillations.\cite{Bosun}

\begin{figure}[h]
\vspace*{2.0cm}
\begin{center}
\includegraphics[width=0.8\textwidth]{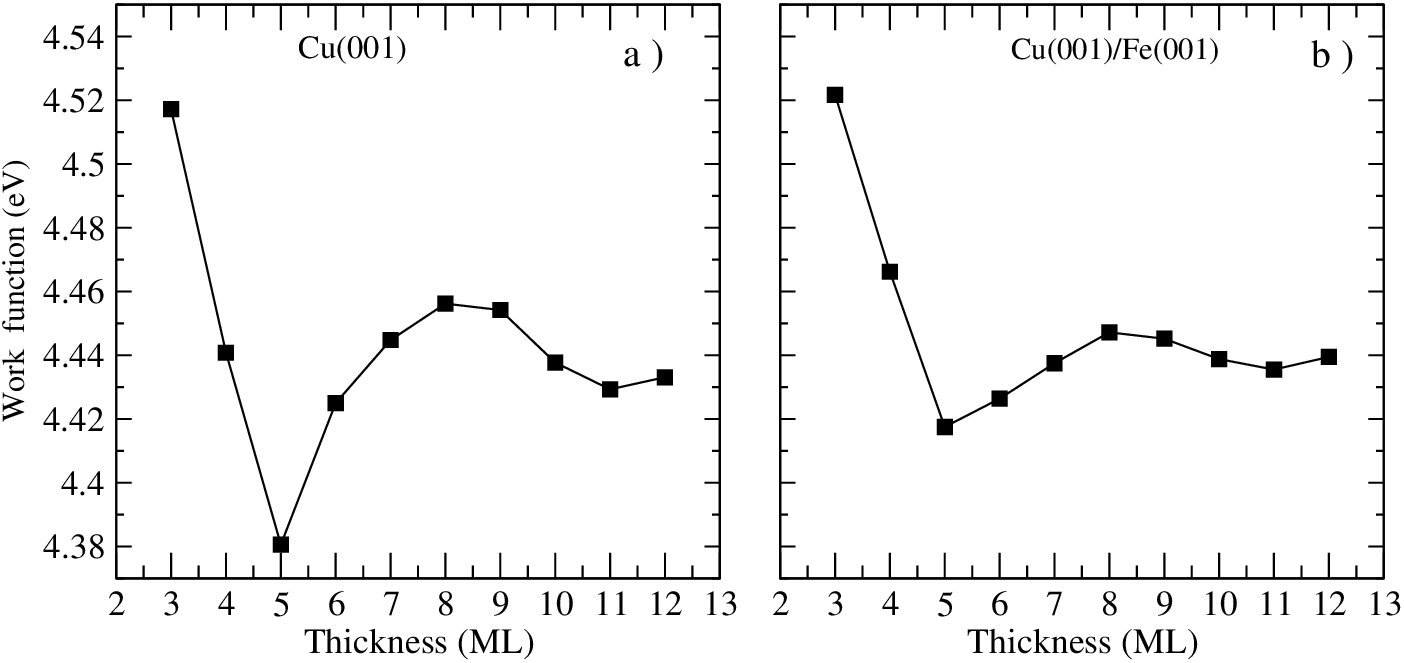}
\caption{\label{wfctus}  Calculated work function of the free-standing 
(left panel) and Fe-supported (righ panel) Cu(001) films, as a function of 
film thickness. }
\end{center}
\end{figure}

In Fig. \ref{wfctus}, we show the theoretical dependence of the work function on 
film thickness, for the free-standing and Fe-supported Cu(001) films. 
The work-function dependence on film thickness is similar for the supported 
and unsupported films,\cite{NoteWF} although the amplitude of the oscillations 
is slightly reduced in the supported case. One can notice periodic cusps, 
corresponding to local minima in the work function at 5 ML and 11 ML, 
which coincide with the thicknesses of the crossing of $E_F$ by a QWS 
at $\bar{\Gamma}$. 
These periodic cusps in the work function, when a QWS at $ \bar{\Gamma} $ 
crosses $E_F$, are consistent with the jellium-model predictions by 
Schulte.\cite{Schulte} We also observe that the work function, in Fig. \ref{wfctus}, 
is largest at 3 ML and shows a local maximum at 8 ML. The calculated 
value at the largest Cu thickness (4.43 eV at 12 ML) is close to the experimental 
work-function value of the Cu(001) surface (4.59 eV).\cite{Gartland} 

An increase/decrease in the work function may be expected to shift the Cu $3d$ 
states to lower/higher energy with respect to the vacuum level and possibly also 
with respect to the CO electronic levels. Therefore one could expect that 
maxima/minima in the work function may lead to a decreased/increased 
interaction (hybridization) between the occupied Cu $3d$ states of the 
surface and the empty CO $2\pi^{*}$ states of the isolated 
molecule, and hence may correspond to minima/maxima in the chemisorption 
energy.\cite{Norskov,Hoff} 
However, comparing the oscillations of the work function (Fig. 
\ref{wfctus}) and of the chemisorption energy of the supported films 
(Fig.\ref{ads1} ), one observes that the local maxima in the chemisorption 
energy (at 4 and 10 ML) are shifted by -1 ML with respect to the minima in the 
work function (at 5 and 11 ML), and that the local minimum of the chemisorption 
energy (at 6 ML) also does not correspond to the local maximum of the work 
function (at 8 ML).  
Furthermore, normal-emission photoelectron spectra\cite{Pears} of Cu(001)  
films indicate a monotonic shift of the Cu $3d$-levels as a function of 
film thickness. The $3d$ centroid moves slightly away from $E_{F} $ with 
increasing thickness, which would tend to monotonically decrease the 
Cu($3d$)-CO($2\pi^{*}$) interaction and monotonically weaken the Cu-CO 
bond.\cite{Norskov,Hoff}  
Hence, for CO on Cu films,  a shift in the $3d$-band energy does not 
appear to be the dominant factor in determining the tend with 
film thickness of the CO chemisorption energy. 

\begin{figure}[h]
\vspace*{1.0cm}
\begin{center}
\includegraphics[width=1.0\textwidth]{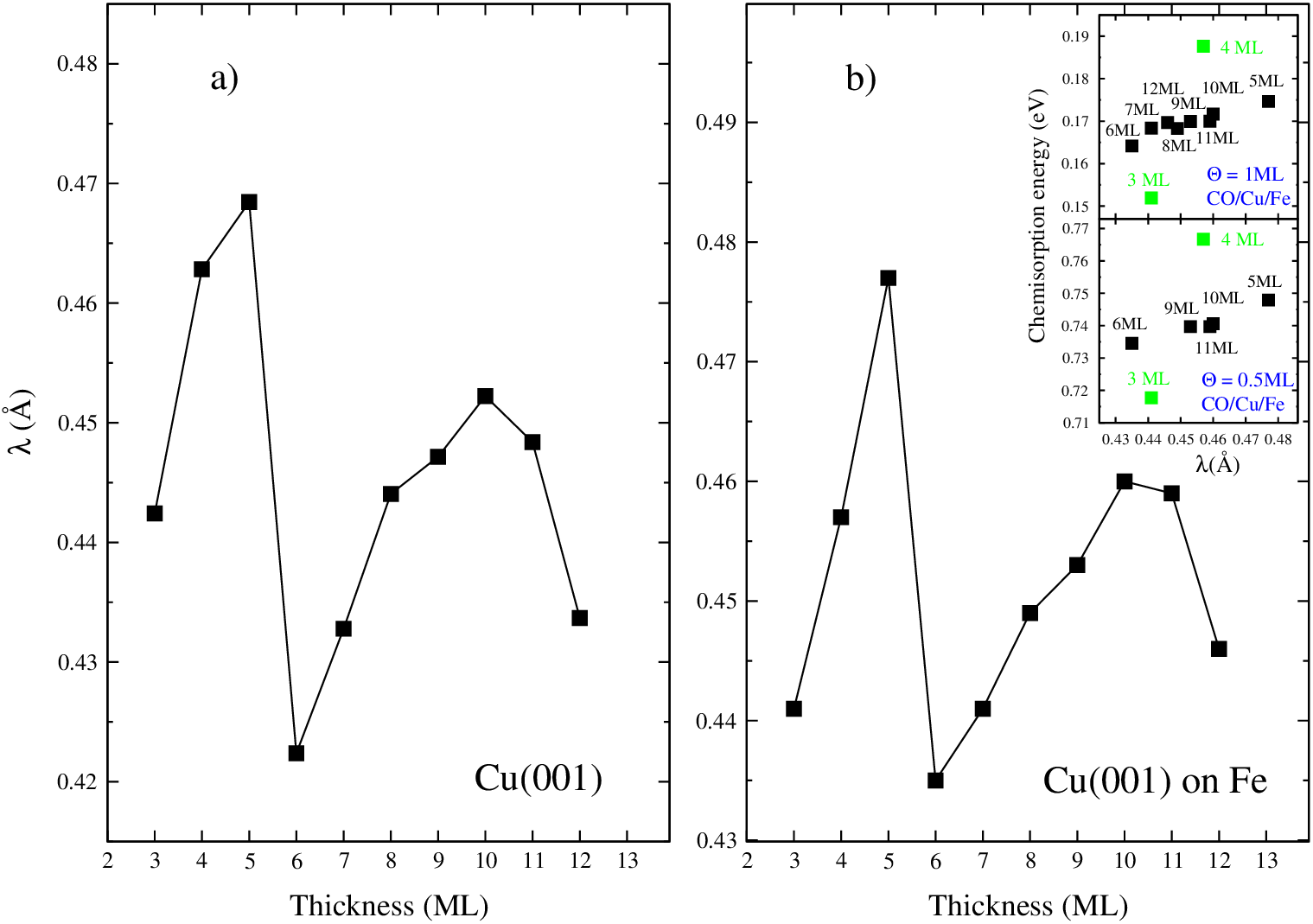}
\caption{\label{lbda}  Calculated decay length in vacuum $\lambda$ of the 
electronic local density of states at the Fermi energy for the free-standing 
(left panel) and Fe-supported (righ panel) Cu(001) films, as a function of 
film thickness. The inset shows the calculated CO chemisorption energy obtained 
for 1 ML (upper panel) and 0.5 ML (lower panel) of CO on the Fe-supported 
Cu(001) films as a function of the decay length $\lambda$ of the 
corresponding Fe-supported Cu film. }
\end{center}
\end{figure}

In Fig. \ref{lbda}, we present the calculated decay length, $ \lambda $, as a
function of Cu film thickness, with and without the Fe substrate. 
The trends are identical in the two cases. The decay length 
exhibits pronounced oscillations, with a first maximum at 5 ML followed by 
a minimum at 6 ML, a gradual increase  to a second maximum at 
10 ML, and a subsequently gradual decrease for thicknesses up to 12 ML. 
We note that this behavior exactly coincide with the experimental trends of 
the desorption temperature in Ref. \onlinecite{Danese}. 
Apart from the local maximum at 4 ML, instead of 5 ML, the  
behavior of the calculated chemisorption energy of the supported films,  
in Fig.\ref{ads1}, also follows the trends of $ \lambda $ in the 
range 5-12 ML (with a local minimum at 6 ML and a local maximum at 
10 ML). In Fig.  \ref{lbda}, we also reported (see inset) the calculated 
CO chemisorption energy obtained for 1 ML (upper panel) and 0.5 ML (lower 
panel) of CO on the Fe-supported Cu(001) films, as a function of the 
decay length $\lambda$ of the supported films. Except for the cases of 
3 and 4 ML ---which are the films for which the chemisorption energy 
critically depends on the substrate (see Fig. \ref{ads1}),
there is a rather clear correlation (linear relation) between the 
calculated chemisorption energy and decay length. One may note that the 
variation in the chemisorption energy is rather small (12 meV) in the 
range 5-12 ML. However, this is consistent with the small variation in the 
desorption temperature reported in Ref. \onlinecite{Danese}. 

The oscillations in $\lambda$ are found to be virtually in phase 
with the periodic crossing of $E_F$ by a QWS at  $ \bar{\Gamma}$. This is 
consistent with the predictions for  $\lambda$ of a particle-in-a-box 
model.\cite{Binggeli2,Binggeli} The model predicts that all states 
$\psi_{n,k_{||}} $ of a given subband $n$ have the same decay 
length $\lambda_n \sim 1/\sqrt{-E_{n}} $, where $E_n$ is the energy of 
the subband state $ n$ at $k_{||} = 0$, measured relative to the vacuum 
level.\cite{Binggeli2,Binggeli} Hence $\lambda$ is dominated by the decay 
length of the states at $E_F$ belonging to the subband whose energy at 
$ \bar{\Gamma} $ is closest to the Fermi level.\cite{NoteLamb} 
With increasing width $L$ of the film, the energy of the highest-occupied 
QWS at $ \bar{\Gamma} $ (of the inverted quantum well) increases with 
respect to $E_F$. Therefore, $\lambda$ first increases as $ \lambda_n 
\sim 1/\sqrt{-E_n}$, until the QWS $n$ at  $\barra{\Gamma}$ crosses 
$E_F$, at which point $\lambda$  decreases to the next value $\lambda_{n+1} 
\sim 1/\sqrt{-E_{n+1}}$ of the highest-occupied QWS at 
$ \bar{\Gamma} $; $\lambda$ then increases again with increasing $L$, 
displaying systematic oscillations with film thickness $L$, as 
observed in Fig. \ref{lbda}.

The effect of $\lambda$ on the chemisorption energy may be understood in 
terms of an increased/decreased overlap, with increased/decreased $\lambda$, 
between the most extended QWS's of the film (on the vacuum side) 
at $E_F$ and the frontier orbitals of the molecule,\cite{Hoff} namely the 
CO $5\sigma$ highest-occupied molecular orbitals (HOMO) and  
$ 2\pi^{*}$ lowest-unoccupied molecular orbitals (LUMO). The 
enhanced/reduced hybridization (a) between the 
unoccupied QWS's just above $E_F$ and the CO $5\sigma$ HOMO 
and (b) between the occupied QWS's of the film at or just below 
$E_F$ and the CO $ 2\pi^{*}$ LUMO are both expected to 
strengthen/weaken the CO bonding to the Cu surface.\cite{Bagus} 
The former interaction is associated with an electronic charge transfer 
from the $ 5\sigma $ orbitals of the molecule to the surface (donation), 
whereas the latter leads to an electronic charge transfer 
from the metal surface to the CO $ 2\pi^{*} $ orbitals (back donation). 
The latter interaction is expected to be the dominant 
one.\cite{Bagus,Norskov,Hoff} 

Hence, for Cu thicknesses larger than 4 ML, the trends we obtain in the 
chemisorption energy can be understood in terms of systematic changes in 
the decay length of the QWS's at $E_F$, which influence the interaction 
between filled/empty QWS's of the film at $E_F^{(+/-)}$ and the LUMO/HOMO   
of the molecule. We note that the calculated and experimental chemisorption 
energy display the same trend as a function of film thickness in the 
range 5-12 ML. The main differences occurs at 4 ML: 
while the experimental chemisorption energy decreases with decreasing 
thickness from 5 to 3 ML, the calculated value has a maximum at 4 ML.  
Such a difference, however, can be expected for the thinnest 
films considered, namely 3 and 4 ML, as the behavior of their 
chemisorption energy sensitively depends on the substrate (see Fig. 4). 
One may therefore also expect the details of the experimental interface 
(e.g., atomic intermixing at the interface) to affect the chemisorption 
energy of these films. 

We note that in our calculations the maximum in the chemisorption energy 
does not occur exactly when a QWS at $ \bar{\Gamma} $  crosses $E_F$, 
but 1 ML before. We believe this is related to the fact that, in the 
calculations, the quantum-well subband with energy closest to $E_F$ at 
$ \bar{\Gamma} $ (see  Fig. \ref{band10}) barely touches the Fermi level 
at 5 ML (and 11 ML). In this situation, although the unoccupied QWS's just 
above $E_F$ may interact with the $5\sigma$ of the molecule, the occupied QWS 
at  $\bar{\Gamma} $ cannot interact, by symmetry, with the CO $2\pi^{*}$ states. 
In fact, based on projections on the atomic states, we find that the QWS's  
near $E_F$ at $ \bar{\Gamma} $ are composed mainly of Cu 4$s$ states, 
with a small component of Cu $3d_{\sigma}$ states (about $10$ \% for the 
$ \bar{\Gamma} $  QWS's within 0.2 eV of the Fermi level).\cite{NoteSym} 
In the region near the Brillouin-zone center, a $3d_{\pi}$ component, 
which can mix with the CO($2\pi^{*}$) orbitals, appears in the QWS's, near 
$E_F$, when going outside $ \bar{\Gamma}$, e.g., along the $ \bar{\Gamma} 
- \bar{\rm M}$ direction. The $3d_{\pi}$ component increases linearly 
with $ k_{||}$ (reaching, e.g., $\sim 10$ \% at 1/5 of the $ \bar{\Gamma} -  
\bar{\rm M}$ distance for the 10 ML QWS with subband energy at 
$ k_{||} = 0$ closest to $E_F$ in Fig.\ref{band10}).\cite{NoteSym} 
In order thus for the QWS's with the largest possible decay length at 
$E_F$  to mix with the $2\pi^{*}$ states, a small, but non-vanishing 
fraction of the corresponding quantum-well subband should be occupied. 
This corresponds to an optimal situation in which the highest occupied 
QWS at $ \bar{\Gamma}$ is closest to, but not yet crossing $E_F$. In 
our calculations, this situation corresponds to the 4- and 
10- ML cases, which in fact yield  maxima in the calculated 
chemisorption energy. 

\section{Summary and conclusions}\label{sec:Summary} 

We have performed first-principles calculations, based on DFT, to study the
modulations with film thickness in the chemisorption energy of CO 
molecules on Cu(001) films. We have examined Fe-supported and free-standing 
Cu(001) films with thicknesses in the range 3-12 ML. The presence of 
the substrate has an important influence on the chemsiorption energy for 
thicknesses of 4 ML and below. The calculated CO chemisorption energy 
of the supported films displays systematic oscillations, as a function of 
film thickness, with a periodicity corresponding to that of quantum-well 
states at $\bar{\Gamma} $ crossing the Fermi energy. These oscillations in 
the chemisorption energy are understood in terms of periodic modulations 
of the decay length in vacuum of the quantum-well states at the Fermi energy. 
These modulations of the decay length are expected to influence the interaction 
of the quantum-well states with the frontier orbitals of the molecule. 
Contrary to previous suggestions, we find that the 
actual oscillations with film thickness of the density of states per atom of 
the films at the Fermi energy cannot account for the observed modulations 
of the chemisorption properties.

\begin{acknowledgments}

One of us (LM) gratefully acknowledges support by the ICTP through the OEA-AC-71 
grant and by the ICTP-IAEA Sandwich Training Educational Programme (STEP). 
We also thank M. Altarelli for useful discussions. The computations have been 
performed on the IBM sp6 computer at CINECA. 

\end{acknowledgments}


\begin{thebibliography}{99}
\bibitem{NoteFe} The theoretical value of the Fe lattice constant is 
3.66 \AA\@ (-0.5 \% lattice mismatch); we considered ferromagnetic Fe.  
For 5 ML of fcc-Fe on Cu(001), we performed test calculations for 
various collinear spin configurations of the Fe layers, and find 
the ferromagnetic configuration to be the lowest in energy 
(followed by the $\uparrow \uparrow \downarrow \uparrow \uparrow$ 
configuration, which is about 0.2 eV higher in energy per flipped  
Fe spin).
\bibitem{NoteDFT} We performed calculations for CO on the Fe-supported 
Cu(001)-(1x1) films using a corrected GGA functional, which modifies the 
CO gap. As in Ref. \onlinecite{Kresse03}, we used a GGA + U  type of 
approach with U$_{CO}$ = 0.75 eV.  We find that this rigidly shifts the 
chemisorption energy, but does not affect (within 0.1 meV) the   
variations of the chemisorption energy with film thickness.
\bibitem{NoteRelax} We note that relaxation of the interlayer spacing 
is found to modify the peak positions of the QWS's near  $E_{F} $ by less 
than 0.1 eV and has no influence on the sequence of crossing.
\bibitem{NoteWF} We note that our results for the work function of the 
Cu(001) films display more regular/systematic oscillations than  
found for the free-standing films in Ref.~\onlinecite{Bosun}. We relate 
this difference to the smaller k-point grid of (11,11,1) used for the 
work-function calculations in Ref.~\onlinecite{Bosun}.
\bibitem{NoteLamb} In addition to displaying the correct trend (i.e., a 
correlation with the chemisorption energy) in the range 5-12 ML, the decay 
length is a much more robust parameter than the DOS in terms of its energy 
dependence around the Fermi energy. For a given film, because the 
decay length is the same for all states belonging to a subband $n$, 
$\lambda(E)$ is a constant when moving around the Fermi energy and changes 
abruptly only when $E$ crosses a QWS at $\bar{\Gamma}$.
\end{thebibliography}
\end{document}